\documentclass[sn-mathphys,Numbered]{sn-jnl}
\usepackage{svg}%
\usepackage{graphicx}%
\usepackage{multirow}%
\usepackage{amsmath,amssymb,amsfonts}%
\usepackage{amsthm}%
\usepackage{longtable}
\usepackage{mathrsfs}%
\usepackage[title]{appendix}%
\usepackage{tabularx,ragged2e}
\usepackage{newtxtext} 

\usepackage{xcolor}%
\usepackage{textcomp}%
\usepackage{hyperref}
\usepackage{manyfoot}%
\usepackage{booktabs}
\usepackage{algorithm}%
\usepackage{algorithmicx}%
\usepackage{algpseudocode}%
\usepackage{listings}%
\usepackage{array}
\usepackage{caption}%
\usepackage{subcaption}%
\usepackage{geometry}

\usepackage{rotating} 
\usepackage{changepage}
\usepackage{tabularx}   
\usepackage{lscape} 

\begin{document}
\title{TMAP: A Threat Modeling and Attack Path Analysis Framework for Industrial IoT Systems (A Case Study of IoM and IoP)}

 \author*[1]{\fnm{Kumar} \sur{Saurabh}}\email{pwc2017001@iiita.ac.in}
\equalcont

\author[1]{\fnm{Deepak } \sur{Gajjala}}\email{iib2019024@iiita.ac.in}
\equalcont

\author[1]{\fnm{Krishna} \sur{Kaipa}}\email{iit2019203@iiita.ac.in}
\equalcont{These authors contributed equally to this work.}

\author[1]{\fnm{Ranjana} \sur{Vyas}}\email{ranjana@iiita.ac.in}
\author[1]{\fnm{O.P.} \sur{Vyas}}\email{opvyas@iiita.ac.in}
\author[2]{\fnm{Rahamatullah} \sur{Khondoker}}\email{rahamatullah.khondoker@mnd.thm.de}
\affil[1]{\orgdiv{Department of IT}, \orgname{Institute of Information Technology, Allahabad, India}, \orgaddress{ \city{Prayagraj}, \postcode{211012}, \state{UP}, \country{India}}}

\affil[2]{\orgdiv{Department of Business Informatics}, \orgname{THM University of Applied Sciences}, \orgaddress{\city{Friedberg}, \postcode{61169}, \state{Hessen}, \country{Germany}}}
              

\abstract {Industrial cyber-physical systems (ICPS) are gradually integrating information technology and automating industrial processes, leading systems to become more vulnerable to malicious actors. Thus, to deploy secure Industrial Control and Production Systems (ICPS) in smart factories, cyber threats and risks must be addressed. To identify all possible threats, “Threat Modeling” is a promising solution. Despite the existence of numerous methodological solutions for threat modeling in cyber-physical systems (CPS), current approaches are ad hoc and inefficient in providing clear insights to researchers and organizations involved in IIoT technologies. These approaches lack a comprehensive analysis of cyber threats and fail to facilitate effective path analysis across the ICPS lifecycle, incorporating smart manufacturing technologies and tools. To address these gaps, a novel quantitative threat modeling approach is proposed, aiming to identify probable attack vectors, assess the path of attacks, and evaluate the magnitude of each vector. This paper also explains the execution of the proposed approach with two case studies, namely the industrial manufacturing line, i.e., the Internet of Manufacturing (IoM), and the power and industry, i.e., the Internet of Production (IoP).}


\maketitle
\section{Introduction}
Technology-driven Industry 4.0\footnote{Industry 4.0 is all about creating "smart" and interconnected production systems that can sense and predict. This allows them to make on-the-spot decisions that greatly enhance production processes.\cite{XU2021530}\cite{2022}}  and value-driven Industry 5.0\footnote{Industry 5.0 represents a new era where robots and smart machines team up with people, prioritizing resilience and sustainability. It takes a step further than Industry 4.0 by putting humans, the environment, and social values at the core of its approach.\cite{XU2021530}\cite{2022}} are continuously revolutionizing industries to make manufacturing and production processes smarter, resulting in notable challenges for industrial cyber security, as the threat landscape is constantly changing with the adoption of new technologies. Enabling various technologies to integrate and operate
seamlessly as a unified system can typically enhance operational efficiency, minimize errors, lower expenses, optimize
workflows, and confer competitive advantages to businesses.
Industrial use of the Internet of Things is an emerging field
where industries combine smart sensors and actuators with
existing machinery. It focuses on intercommunication between
machines, automation of repetitive tasks, machine learning,
and big data analytics. Operational efficiency has increased
drastically due to the advent of Internet of Things (IoT) technology and continues
to change how industries innovate and grow. Industrial IoT
is also different from consumer IoT in the sense that it is a
combination of Information Technology (IT) and Operational
Technology (OT). "IT" refers to early warning systems, insights
from data, machine learning, data aggregation, and data pro-
cessing, while "OT" involves Industrial Control Systems (ICSs),
Human-Machine Interfaces (HMIs), Programmable Logic Con-
trollers (PLCs), Supervisory Control and Data Acquisition (SCADA) systems, and Distributed Control Systems (DCSs). This
convergence of IT and OT enables industries to use data collected from operating machines to provide real-time insights
for optimizing industrial processes, transforming traditional
industries into smart industries. These integrations are nowadays also coined as the Internet of Manufacturing\footnote{What differentiates "Production" from "Manufacturing" and vice versa, the Manufacturing is defined as a sequence of interrelated activities and processes comprising the design, material selection, planning, production, quality assurance, management, and marketing of discrete consumer and durable goods\cite{10.1145/3502265}} (IoM)\cite{doi:10.1080/24725854.2018.1555383} for smart manufacturing and the Internet of Production\footnote{The definition of production is "the process through which raw materials are transformed into completed goods"\cite{10.1145/3502265}} (IoP)\cite{8780276} for smart production.
\par Industrial systems comprise several layers of complex machinery, and the integration of IoT presents a new set of
challenges\cite{JBAIR2022103611}, including ensuring safe operation and resilience
to attacks from malicious actors\cite{9817297, 9854947}. Infrastructure security is
crucial in critical industries such as energy, chemical, and gas
manufacturing, and healthcare because an attack on the system
could endanger countless lives and cause significant financial
losses to stakeholders. The complexity of industrial IoT systems directly leads to a large attack surface\cite{9817245}, meaning that the number of attacks possible is substantial\cite{maggi2020attacks}. Understanding where
attacks can emerge while also taking the severity of the
attack into account is a fundamental way to secure an industrial
IoT system\cite{10039777},\cite{magar2016state}.
\par In this paper, a quantitative threat modeling and path analysis method for industrial IoT systems has been proposed. In this work, two case studies have been
taken for the implementation of the proposed model, one being
the power industry (IoP) and the other being the manufacturing industry (IoM). The advantage of taking a generalized approach is that
it gives you the flexibility to apply the framework to several
different industries. This study offers significant contributions
to researchers and organizations involved in the Industrial Internet of Things (IIoT)
technologies. The key research contribution of this manuscript is as follows:
\begin{itemize}
\item \textbf{Proposed Quantitative Threat Modeling and Path Analysis Method:} The paper introduces a new method for conducting quantitative threat modeling and path analysis specifically designed for industrial IoT systems. This approach provides a systematic and structured way to assess and analyze cyber threats in such environments.
\item \textbf{Case Studies in Power and Manufacturing Industries:} The paper includes two case studies in the power industry (IoP) and the manufacturing industry (IoM) to demonstrate the practical implementation of the proposed model. By applying the framework to these different industries, the research showcases its versatility and applicability.
\item \textbf{Generalized Approach:} One notable advantage of the proposed model is its generalized approach, allowing it to be applied across various industries. This flexibility is beneficial for researchers and organizations involved in the Industrial Internet of Things (IIoT) technologies, as it offers a framework that can be adapted to different industrial sectors.
\item \textbf{Comprehensive Analysis of Cyber Threats:} The primary focus of the research is to provide a thorough analysis of cyber threats specifically targeted at industrial scenarios. Unlike other works in IIoT attack modeling, this study aims to offer an up-to-date and valid reference framework that captures the rapidly evolving industrial environment.
.
\end{itemize}
Overall, the paper contributes to the field of Industrial IoT by providing a quantitative threat modeling and path analysis method, demonstrating its effectiveness through case studies, offering a generalized approach applicable to various industries, and providing a comprehensive analysis of cyber threats specific to industrial equipment.
\par In order to put our research into context, the background and motivation of our study have been addressed in sections I \& II. The remainder of the article is structured as follows: Section III outlines the proposed methodology,
in which an attack path assessment framework has been pro-
posed. Section IV discusses the implementation
of the proposed threat modeling and attack path analysis by
using two case studies. while Section V concludes the paper
and discusses future research directions.

\section{Metasurvey}
To conduct a review of the state-of-the-art methodologies and to explore how cyber security and threat modeling are relevant to industrial IoT threat modeling and attack path analysis, the literature study work has been performed. The review specifically examines cyber security standards and the latest methods for modeling threats in ICS.
\subsection{Related Works}
Threat assessment in industrial IoT is a growing research topic with several notable works. The closest and most recent is a study by Jbair et al.\cite{JBAIR2022103611} which looked at threat analysis in the smart manufacturing industry. The study proposed a method of threat modeling and risk analysis in systems specifically for the smart manufacturing industry. Their threat modeling process included noteworthy elements such as a factor of asset importance, albeit with a non-standard scoring system. Lorenzo et al.\cite{10.1145/3381038} surveyed all available protocols in industrial IoT networks and used a CVSS to quantify the threat each protocol poses. The study also presents a framework to characterize the severity of each protocol in the IIoT environment.
\par To ensure that this study covered all possible cyber threat avenues when modeling for industrial IoT systems, a thorough analysis of all possible cyber threats in IIoT was essential. Hassija et al.\cite{10.1109/ACCESS.2019.2924045} achieved the former requirement by describing various security issues present in an IoT application at different levels, including the sensing layer, network layer, middleware layer, gateways, and application layer. The study also presents an overview of upcoming cybersecurity threats in fields such as blockchain and cloud computing. A literature survey into cybersecurity issues in industrial IoT by Jayalakshmi et al.\cite{10.1109/ACCESS.2021.3057766} took a deeper look into the architecture of an industrial IoT system and the solutions proposed for attack detection. These studies served as a catalog of current security issues and not as a threat model for IoT systems. 
\par The modeling done by Falco et al.\cite{10.1109/JIOT.2018.2822842} was a specific approach focusing exclusively on SCADA systems to draw a data-driven organizational schema that could help researchers to recognize security breaches that are crucial to our operations, so they presented a framework to prioritize the modeled threats using the CVSS scoring system. To ensure that researchers choose the necessary threat modeling approach, Nataliya Shevchenko et al. \cite{12_methods} comprehensively explained the twelve threat modeling methods, namely STRIDE (Spoofing, Tampering, Repudiation, Information Disclosure, Denial of Service (DoS), and Elevation of Privilege) and their associated derivations. PASTA 
(Process for Attack Simulation and Threat Analysis), LINDDUN(Linkability, Identifiability, Non-Repudiation, Detectability, Disclosure of Information, Unawareness, and Non-Compliance), CVSS(Common Vulnerability Scoring System), Attack Trees, Persona Non-Grata, Security Cards, HTMM (Hybrid Threat Modeling Method), Quantitative Threat Modeling Method, Trike, VAST (Visual, Agile, and Simple Threat) Modeling, OCTAVE (Operationally Critical Threat, Asset, and Vulnerability Evaluation), that came from various articles, and they focus on diverse functions of the process. Each threat modeling method is special and is not suggested over others; the selection of which methods to use should be based on the project's requirements and its considerations\cite{12_methods}. These studies have been found useful as a checklist for covering all security issues for creating a threat model for an industrial IoT system.
\par The comprehensive analysis encompasses a diverse array of methodologies and frameworks employed in the field of threat modeling and risk analysis within the context of industrial Internet of Things (IoT) systems.  Certain studies have been dedicated to examining particular industries or protocols within the realm of industrial Internet of Things (IoT), aiming to gain a deeper understanding of cybersecurity challenges within these specific contexts. Conversely, other research endeavors adopt a broader perspective, seeking to offer a comprehensive overview of the various cybersecurity issues that arise in the industrial IoT landscape as a whole. The scoring systems employed in various studies exhibit considerable variation, with certain studies adopting the Common Vulnerability Scoring System (CVSS), while others opt for non-standard scoring systems predicated on the significance of assets. The aforementioned studies provide significant insights pertaining to the establishment of a comprehensive threat model for an industrial Internet of Things (IoT) system. The process involves the identification of security concerns across various layers, the prioritization of risks that have been modeled, and the selection of the most suitable threat modeling methodologies in accordance with the project's requirements and considerations.

\begin{table*}[htbp]
\centering
\caption{Summary of Literature Survey on Threat Assessment in Industrial IoT Systems}
\label{tab:literature-survey}
\begin{tabular}{p{1.5cm}p{2cm}p{2cm}p{2cm}p{2cm}p{3cm}}
\hline
Study & Focus & Methodology & Frame-work & Scoring-System & Key Findings \\ 
\hline
Jbair et al.~\cite{JBAIR2022103611} & Threat analysis in smart manufacturing industry & Threat modeling and risk analysis & N/A & Non-standard scoring system based on asset importance & Proposed a method for threat modeling and risk analysis specifically for smart manufacturing industry \\ 
\hline
Lorenzo et al.~\cite{10.1145/3381038} & Survey of available protocols in industrial IoT networks & Common vulnerability scoring system (CVSS) & Framework to characterize the severity of each protocol & CVSS & Framework to characterize the severity of each protocol in the IIoT environment \\ 
\hline
Hassija et al.~\cite{10.1109/ACCESS.2019.2924045} & Identification of security issues present in IoT applications at different levels & N/A & N/A & N/A & Identified various security issues present in an IoT application at different levels \\
\hline
Jayala-kshmi et al.~\cite{10.1109/ACCESS.2021.3057766} & Literature survey into cybersecurity issues in industrial IoT & N/A & N/A & N/A & Cataloged current security issues in industrial IoT systems \\
\hline
Falco et al.~\cite{10.1109/JIOT.2018.2822842} & Modeling of threats in SCADA systems & Data-driven organizational schema & Framework to prioritize modeled threats & CVSS & Proposed a framework to prioritize modeled threats in SCADA systems \\
\hline
Nataliya-Shevchenko et al.~\cite{12_methods} & Comprehensive explanation of twelve threat modeling methods & N/A & N/A & N/A & Comprehensive explanation of twelve threat modeling methods and their associated derivations\\
\hline
\textbf{Proposed Work (TMAP)} & Asset-centric versatile and generalized approach that can be applied across diverse industries for Threat analysis & Quantitative Threat Modeling and Attack Path
Analysis using Purdue Model &  Framework for cyber threats specific to the Manufacturing and production Industry. & CVSS & Proposes a robust framework for quantitative threat modeling and path analysis, validated through insightful case studies.\\
\hline
\end{tabular}
\end{table*}

\subsection{Research Gap} 

In contrast to the existing research gap, there is a plethora of research on threat modeling and detection, there is a conspicuous absence of a standardized, step-by-step process in the literature for identifying, assessing, and prioritizing threats, especially in the context of IIoT, none of the models and methodologies presented in the findings are tailored for specific industries. While these offer valuable insights for particular sectors, their applicability becomes limited when faced with the diverse and evolving landscape of the Internet of Things across various industries.

\par Here are the potential research gaps identified in the related work section:

1. \textbf{Lack of a Standardized, Step-by-Step Process:} There is a research gap regarding the absence of a standardized, step-by-step process for identifying, assessing, and prioritizing threats in the context of Industrial IoT. While there is existing research on threat modeling and detection, the absence of a structured methodology for addressing threats, especially in specific industries, is highlighted as a gap.

2. \textbf{Limited Applicability to Diverse Industries:} The existing models and methodologies are criticized for their limited applicability when faced with the diverse and evolving landscape of the Internet of Things across various industries. The research does not tailor these models for specific industries, and this lack of adaptability is identified as a research gap.

3. \textbf{Absence of Metrics and Criteria:} The paragraph emphasizes the need for metrics and criteria for assessing and prioritizing threats. While threat identification is mentioned, the absence of specific metrics and criteria for evaluating and ranking these threats is considered a gap in the existing research.

4. \textbf{Industry-Agnostic Approach:} The research introduces an industry-agnostic approach, highlighting the need for versatile and adaptable methodologies. The gap is that many existing methodologies are not versatile enough to be relevant across a wide spectrum of industries, limiting their utility and impact.

5. \textbf{Case Studies for Validation:} There is a research gap in terms of case studies for validation and practical application.

\subsection{State of The Art- Architectural Setup Methodologies } 

\par The present section endeavors to investigate the indispensability of an architectural configuration for the purpose of threat modeling. The central inquiry being examined pertains to the necessity of implementing such a configuration. The first step in the process of threat modeling entails the development of a customized architectural framework that is specifically designed to address the unique requirements and characteristics of the industry being examined. It is crucial to possess a comprehensive comprehension of the diverse components within the industry and the intricate connections that exist between them. Through the execution of this process, individuals are able to proficiently identify potential avenues through which attacks may occur. This enables them to subsequently devise and implement suitable strategies and safeguards to mitigate and neutralize these threats. The present study focuses on the comprehensive examination of the various components involved in the configuration process. Specifically, it entails the meticulous identification of the distinct data categories that are intended to be shared within the system. Furthermore, it involves the systematic classification of the diverse user types that will be granted access to the aforementioned data. Lastly, it encompasses the rigorous contextual analysis of the environment in which these users will actively interact with the data.
After establishing the architectural framework, one can initiate the process of threat identification by assessing potential attack vectors. The task at hand pertains to the classification of multiple kinds of threats and their respective ramifications. You'll also want to think about how you can prevent such threats from occurring in the first place. Finally, you can consider any security measures that you can put in place.
\par There are numerous architectural setup methodologies have been proposed by various research groups and organizations. Explaining all of those architectures is not in the scope of this manuscript but to have a better understanding of the architectural setup some popular architectural models are tabulated in Table~\ref{label} in this manuscript.

\begin{table*}[!htbp]
\caption{Summary of Literature Survey of State of The Art- Architectural Setup Methodologies}
\label{label}
\centering
\begin{tabular}{p{.7cm}p{1.7cm}p{3.7cm}p{1.3cm}p{6cm}}
\hline
\textbf{S.No.} & \textbf{Architecture }&\textbf{Description }& \textbf{Industry} & \textbf{Key Features }\\
\hline
1  & Purdue Model~\cite{BOYES20181} & A hierarchical model for designing and integrating industrial control systems & Manufac-turing & Five levels of the model include Enterprise, Site, Area, Cell, and Process - Each level is responsible for different tasks and functionalities - Provides a structured approach for connecting different devices and systems within a manufacturing plant \\
\hline
2  & ANSI/ISA-95~\cite{doi:10.1177/1687814018784192} & A standard for enterprise and control system integration in the manufacturing industry & Manufac-turing  &  Defines a standardized architecture for integrating control systems, manufacturing operations management, and enterprise systems - Provides a common language and framework for communication between systems  \\
\hline
3  & IEC 62443~\cite{10.1145/3339252.3341481} & A standard for industrial network and system security & Any & Provides a framework for implementing cybersecurity measures in industrial control systems - Includes a comprehensive set of security requirements, such as network segmentation, access control, and incident response planning - Designed to be flexible and scalable, allowing organizations to tailor their security approach to their specific needs  \\
\hline
4  & OSA-CBM~\cite{amaya2009open} & A framework for interoperability and data exchange between industrial automation systems & Any & Based on a service-oriented architecture (SOA) - Provides a standardized approach for exchanging data and information between different systems and devices - Enables real-time communication and data analysis across multiple systems and devices\\
\hline
5  & Zachman~\cite{Urbaczewski_Mrdalj} & A framework for organizing and classifying architectural artifacts in an enterprise & Any & Organizes architectural artifacts into a matrix based on six perspectives: Who, What, Where, When, Why, and How - Provides a common language and structure for communication between stakeholders in an organization - Enables a holistic view of an organization's architecture \\
\hline
6  & TOGAF~\cite{Urbaczewski_Mrdalj}  & A framework for enterprise architecture development and management & Any & Provides a standardized approach for developing and managing enterprise architecture - Includes a comprehensive set of tools and techniques for architecture development, such as a reference architecture, architecture content framework, and architecture maturity model.   \\
\hline
7  & DoDAF~\cite{Urbaczewski_Mrdalj} & A framework for enterprise architecture development and management in the US Department of Defense & Defense & Provides a standardized approach for developing and managing enterprise architecture within the Department of Defense - Includes a comprehensive set of tools and techniques for architecture development, such as a capability view, operational view, and systems view - Enables the Department of Defense to align its strategic goals with its IT investments \\
\hline
\end{tabular}
\end{table*}
\begin{table}[!htbp]

\centering

\begin{tabular}{p{.7cm}p{1.7cm}p{3.7cm}p{1.3cm}p{6cm}}
\hline
\textbf{S.No.} & \textbf{Architecture }&\textbf{Description }& \textbf{Industry} & \textbf{Key Features }\\
\hline
8  & FEAF~\cite{Urbaczewski_Mrdalj} & A framework for enterprise architecture development and management in the US Federal Government & Govt. & Provides a standardized approach for developing and managing enterprise architecture within the US Federal Government - Includes a comprehensive set of tools and techniques for architecture development, such as a reference model, data reference model, and service reference model.\\

\hline
9  & Gartner~\cite{6702772} & A framework for enterprise architecture development and management developed by Gartner, Inc. & Any & Provides a comprehensive approach for developing and managing enterprise architecture, with a focus on IT governance and strategy - Includes a set of best practices and tools for\\
\hline
10  & IIRA~\cite{NAKAGAWA2021107241} & An architectural framework with guidelines, methods, and templates to maximize value creation in CPS-based IIoT systems & Any & A comprehensive and adaptable framework for developing CPS-based IIoT systems is offered by the Industrial Internet Reference Architecture (IIRA). Focusing on end-user goals, functional components, implementation assistance, cross-cutting concerns, and fostering collaboration and integration throughout the IIoT ecosystem are some of its primary aspects.
\\
\hline
11  & RAMI 4.0~\cite{NAKAGAWA2021107241} & Provides guidelines on implementing flexible SOA to enable key aspects of CPS-based I4.0 & Any & RAMI 4.0 provides a flexible, standardized foundation for Industry 4.0 applications. The hierarchical organization, life cycle and value stream management, standardized interfaces, flexible SOA, and security make it perfect for establishing efficient and effective industrial systems.
\\
\hline

12  & IMSA~\cite{NAKAGAWA2021107241}  &  An architecture made up of system lifecycle, system hierarchy, and intelligent functions dimensions. & Any & The IMSA design optimizes industrial processes.  The system lifecycle involves the design, production, logistics, etc. processes, while the system hierarchy involves cooperating components from the low equipment level to the highest enterprise level. The intelligent functions involve interoperability frameworks and resources for interconnections with other architectures or patterning of new business models.
\\
\hline

13  & 5C~\cite{doi:10.1177/1687814018784192} & Involves 5 layers of the architecture, namely Connection, Conversion, Cyber, Cognition, and Configure & Any & The 5C IIoT architecture emphasizes a cyclical data flow, moving from the physical interfaces to the cyber interfaces and back again, creating a continuous feedback loop that allows for real-time monitoring and optimization of industrial processes.\\
\hline
14  & 8C~\cite{doi:10.1177/1687814018784192} & Integrates three non-stacked layers called "facets" that horizontally integrate the 5C layers & Any & The manufacturing industry's 8C architecture, based on ISA-95 and 5C architectures, uses three non-stacked "facets" to improve Cyber-Physical Systems (CPS). Parallel facets horizontally combine the 5C layers.
\\
\hline
\end{tabular}
\end{table}

\subsection{Threat Modeling Methodologies}
The process of identifying, enumerating, classifying, and mitigating possible dangers is referred to as "threat modeling." It is a preventative method that is utilized to gain an understanding of how various dangers and assaults could be carried out\cite{Welekwe_2021}. The objective of threat modeling is to supply security teams with a methodical study of what countermeasures need to be developed, taking into account the characteristics of the asset, the attack vectors that are most likely to be used, and the assets that are most desired by an adversary. Threat modeling provides answers to questions such as:
\begin{itemize}
    \item "Where it is most vulnerable to attack?" 
    \item "What threats if carried out may result in a greater impact?" \item "What countermeasures are required to safeguard against these threats?"
    \end{itemize}
\par In general, there are three basic approaches to threat modeling: software-centric, attacker-centric, and asset-centric.
\par\textbf{1. Software-Centric Approach:}The software-centric approach prioritizes the examination of the software system in order to detect possible security risks and vulnerabilities. It requires having an understanding of the architecture, design, and specifics of the execution of the system. The objective of this methodology is to identify vulnerabilities present in the software code, system configuration, and the overarching security stance of the system. Commonly, it encompasses tasks such as scrutinizing code, conducting security assessments, and evaluating the constituents, connections, and information transmissions of the system.
\par \textbf{2. Attacker-Centric Approach:}
The approach that is centered on attackers takes into account the viewpoint of possible attackers and their tactics. The process entails the identification of possible threats through the examination of the motives, abilities, and potential avenues of attack of adversaries. This methodology facilitates comprehension of the methodologies that malevolent actors may employ to capitalize on security weaknesses. The approach of considering the attacker's perspective entails directing attention toward the identification of possible vulnerabilities that may be leveraged to undermine the security of the system. This methodology frequently encompasses methodologies such as the acquisition of threat intelligence, conducting penetration testing, and emulating plausible attack scenarios.
\par \textbf{3. Asset-Centric Approach:}
The proposed work is based on the attack-centric approach which is centered on the identification and safeguarding of essential assets within a given system. The process entails the classification and ranking of assets according to their significance, sensitivity, and worth to the enterprise. The objective is to recognize plausible hazards that may have an impact on said resources and subsequently rank the measures taken to alleviate them in order of importance. This methodology facilitates knowledge of the possible effects of a security breach on the vital assets of the enterprise. The process frequently encompasses tasks such as managing an inventory of assets, evaluating potential risks, and selecting suitable security measures to protect the assets that have been identified.

\par It is noteworthy to mention that these methodologies are not fundamentally mutually exclu-
sive, and they can be utilized in conjunction to enhance the comprehensive threat modeling
procedure. Organizations frequently opt for a blend of these methodologies, taking into account
their particular needs and the characteristics of the system or application under inspection. There are various threat modeling approaches that organizations may use in a systematic manner to identify and address possible dangers. The following four approaches are often used:
:

\subsubsection{STRIDE}The STRIDE~\cite{https://doi.org/10.4218/etrij.2021-0181} approach was created by Microsoft and is a commonly employed methodology that offers a systematic framework for conducting threat modeling. Spoofing, Tampering, Repudiation, Information Disclosure, Denial of Service, and Elevation of Privilege are the six STRIDE threats. By considering these categorizations, the STRIDE framework offers a systematic approach to identifying credible threats.

\subsubsection{DREAD}
Microsoft also developed the widely-used DREAD technique~\cite{https://doi.org/10.4218/etrij.2021-0181}, which provides a standardized framework for performing threat modeling. Damage potential, Reproducibility, Exploitability, Affected users, and Discoverability are the five unique risk aspects represented by the DREAD mnemonic. Each component is given a score between 1 and 10 that reflects the degree of risk it poses. Potential risks associated with known threats can be evaluated and prioritized with the help of the DREAD framework.

\subsubsection{OCTAVE} The Operationally Critical Threat, Asset, and Vulnerability Evaluation (OCTAVE) ~\cite{https://doi.org/10.4218/etrij.2021-0181} technique is a systematic approach to identifying and evaluating vulnerabilities and threats to an organization's most important assets. People from different departments or roles within an organization work together in this method. The study methodology consists of three main steps: developing asset-based threat profiles; identifying potential threats; and conducting vulnerability assessments. The OCTAVE methodology takes a methodical approach to evaluating risks, with a spotlight on recognizing and appreciating the value of a business's assets.

\subsubsection{PASTA} The Process for Attack Simulation and Threat Analysis (PASTA)~\cite{https://doi.org/10.4218/etrij.2021-0181} is a risk-centric approach to threat modeling that combines elements from other approaches like STRIDE, attack trees, and risk analysis. There are seven distinct phases in total: planning, detecting, evaluating, reacting, and checking in on progress. The PASTA methodology prioritizes an iterative and adaptable approach, enabling organizations to consistently enhance their threat models by incorporating new information. The primary aim of this endeavor is to recognize plausible dangers, evaluate their possible outcomes and probability, and offer pragmatic suggestions for reducing linked hazards. The aforementioned methodologies function as systematic frameworks that provide direction to entities in executing thorough threat modeling exercises. 

These methodologies serve as structured frameworks that offer guidance to organizations in conducting comprehensive threat modeling exercises. It is noteworthy that entities have the capacity to modify and tailor these methodologies in accordance with their individualized necessities, industry prerequisites, and the intricacy of their systems.

It is imperative to note that threat modeling is a continuous process that necessitates periodic review and modification to accommodate system changes, emerging threats, and evolving security needs. In article~\cite{sym14030549}, the authors compared various threat modeling methodologies and concluded that the STRIDE method is the one used by the vast majority of people.

\subsection{Vulnerability scoring}
Vulnerability scoring systems are managed by a wide range of governmental, commercial, and charity organizations, each of which focuses on a unique component of gauging vulnerability. Bharadwaj R. K. Mantha et al.\cite{inproceedings} found in their study that a Vulnerability scoring system like CERT/CC determines if the Internet infrastructure is in danger and creates a process to analyze the prerequisites for exploiting the vulnerability. However, the SANS vulnerability analysis scale takes into account both default configurations and client/server system vulnerabilities. Microsoft's proprietary scoring methodology, on the other hand, takes into account both the difficulty of exploiting the flaw and its overall impact. However, the systems' applicability to building projects and networks with participants of varied security levels is limited by the assumption that the vulnerability's impact is uniform for everyone and every organization. However, the Common Vulnerability Scoring System (CVSS) assigns numerical scores to vulnerabilities to assess their severity to each vulnerability that has been discovered. Security analysts, industry experts, organizations, and researchers may all speak the same language when assessing vulnerabilities thanks to this widely adopted framework based on relative security levels.
\par The CVSS scoring system is in widespread use, with users ranging from the federal US government to NIST. The CVSS is an industry-standard vulnerability rating methodology. CVSS measures vulnerability and reflects its severity. It contains three groups of metrics: base, temporal, and environmental. It is a common and standard method to assess a threat, assigning a score between 0 and 10, with 10 being the most severe threat and 0 meaning practically no threat (Fig. \ref{cvss_rating}).

\par Modern IIoT-based industries have shifted from manual control to SCADA systems, which give them access to control processes locally or from a remote location and to monitor, gather, and process real-time data. Falco et al.\cite{10.1109/JIOT.2018.2822842} found that for these IIoT networks involving SCADA systems, CVSS metrics are highly insightful and are an accepted industry standard for vulnerability scoring. The study also shows that CVSS assessment is customized to a wide range of industrial applications, meaning this quantitative approach has applicability in various industries\cite{10.1109/JIOT.2018.2822842}.

\begin{figure*}[h]
\centerline{\includegraphics[width=.3\textwidth]{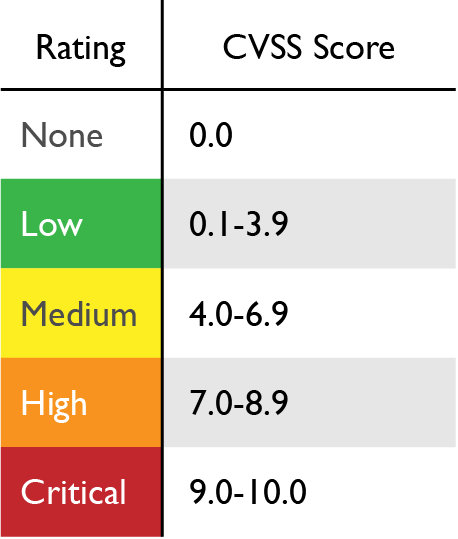}}
    \caption{CVSS V3.1 Rating}
\label{cvss_rating}
\end{figure*}

\section{Proposed Framework}
In Fig.~\ref{Methodology}, An attack path assessment framework has been proposed that takes into account the position of the IIoT device from various viewpoints that are applicable in terms of the device's vulnerabilities towards threats from both physical and cyber perspectives. This procedure can be categorized into the following steps:

\begin{figure*}[h]
  \centering
\includegraphics[width=1.0\textwidth]{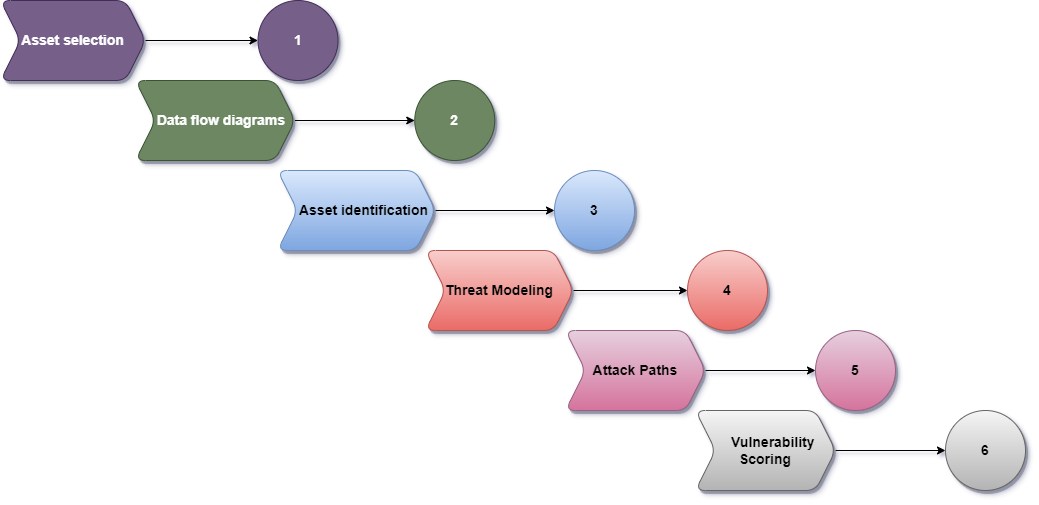}
\caption{Proposed Methodology}
\label{Methodology}
\end{figure*}

\subsection{Architectural setup}
The first step in the threat modeling approach is setting up an architectural setup for a selected set of industries and by setting up the right architectural setup and analyzing potential attack vectors, you can greatly reduce the risk of any potential threats. In the proposed approach, this is done with the help of the Purdue model. The Purdue Model fits better when compared with other popular architectural setup methodologies in Table I. 
All popular models share similarities with the Purdue Model in that they define a hierarchical structure for organizing enterprise information systems. However, they differ in their focus and level of detail, depending on the specific needs of their respective industries. While the Purdue Model is specific to the manufacturing industry, other industries have developed similar reference architecture models to organize and standardize their enterprise information systems.
\par When it comes to whether or not the Purdue model is still accurate and relevant today, that depends on the assets being used. Although it may not be an exact fit for an OT blueprint in an IIoT environment, a three-component architecture may be needed instead of sensors and devices, field/cloud gateways, and a services backend, it's still a great place to start for securing an OT network~\cite{SANS_Institute_2023}.

\subsection{Data flow diagrams}
To assess the threats to an asset, its entries, events, and boundaries could be recognized in the system. It could be achieved by building data flow diagrams (DFDs) that map out the flow of information in the system. It uses varied symbols and routes between each destination. Generally, STRIDE evaluates the system detail design and an accurate DFD shows how successful our STRIDE modeling can be, which is done in the next step. Each connection between entities represents the flow of data and based on this connection, the threats associated with it are also considered.

\subsection{Asset Identification}
The process of asset identification in IIoT, as depicted in the Data Flow Diagram (DFD), entails the identification of various assets that participate in the data flow within the IIoT system.  Assessing the potential security risks associated with each asset and implementing appropriate measures to secure them is a critical step. To gain insights into the assets present in the Industrial Internet of Things (IIoT) through the utilization of Data Flow Diagrams (DFD), a comprehensive analysis of the DFD is imperative. The comprehensive analysis should encompass the identification of the discrete processes, data stores, and external entities that are involved in the transfer of data. Each and every constituent within a given context can be regarded as a valuable asset, warranting a comprehensive evaluation of the associated risks. In the realm of IIoT, the identification of assets via DFD assumes a pivotal role in ensuring the security of IIoT systems. By engaging in the systematic procedure of risk identification and subsequent evaluation for every asset within the IIoT system, it becomes feasible to implement appropriate measures aimed at mitigating the identified risks. This proactive approach serves to bolster the overall security of the IIoT system, thereby minimizing potential vulnerabilities and enhancing its resilience. The identification of assets from a Data Flow Diagram (DFD) entails a systematic approach that can be undertaken by following a series of recommended steps. These steps serve as a guideline to ensure a comprehensive and accurate identification process:
\begin{enumerate}
  
  \item To begin, examine the system's context-level diagram to determine the external actors in the scenario. They could be individuals, other systems, or even whole corporations.

  \item Find out what steps are taken inside the system. Each process is a representation of a unique system function.

  \item Follow the trail of information as it moves from one procedure to another and beyond. This will aid in determining the nature of the information being sent and the appropriate security measures to take.

  \item Look for data stores within the system. These are places
where data is stored and can be considered assets.
  \item Identify any connections or interfaces between the system and other systems or networks. These connections
could be potential attack vectors and should be secured
appropriately.
\end{enumerate}
By following these steps, you should be able to identify the
assets within a system and take appropriate steps to protect
them.
\subsection{Threat Modeling}
STRIDE is a popular threat modeling paradigm that is an acronym for the various threat categories it encompasses: Spoofing, Tampering, Repudiation, Information Disclosure, Denial of Service, and Elevation of Privilege. STRIDE is considered the most mature threat modeling methodology. The reasoning behind choosing Stride in place of other alternatives is due to its low false positive rate and comprehensive aggregation of all possible attacks (12 methods). The downside of using the STRIDE method is the large number of modeled threats is obtained. This is later taken care of by threat rating, which facilitates the prioritization of the attack vectors with a higher rating. Table ~\ref{tab1} shows what each attack category means.

\begin{table}[h]
\label{table1}
\caption{Elements of STRIDE}
\begin{tabular}{|c|c|}
\hline
\textbf{Threat}&\textbf{Threat Definition} \\ 
\hline
Spoofing Identity & Disguising the real identify to appear as \\ & trusted source \\
\hline
Tampering with data & Modifying data without permission \\ 
\hline
Repudiation & Denying taking part in a transaction falsely  \\ 
\hline
Information Disclosure & Revealing sensitive data to unauthorized \\ & entities \\ 
\hline
Denial of Service & Denying access to resource or data \\ 
\hline
Elevation of privilege & Gaining unauthorized access of elevated rights \\ 
\hline
\end{tabular}
\label{tab1}
\end{table}

"Microsoft Threat Modeling Tool"\cite{jegeib} has been used to identify potential security threats to software applications, systems, or networks. It is used to design, analyze, and refine security models for applications and industrial networks. This tool uses a systematic approach (STRIDE) to model threats, which has common categories of threats that can be used to analyze and mitigate security risks. Using the Microsoft Threat Modeling Tool\cite{jegeib} , DFD is created, and this tool, in turn, models and generates a report of potential security vulnerabilities\cite{jegeib}.

\subsection{Using attack vectors to create attack paths}
Attack paths are a useful way to diagrammatically model potential attack vectors. It depicts attacks on a system in path form. The root of the path is represented by a goal for the attack, and its branches are ways to achieve that goal. When the threat analysis of a system is performed, it has several threat-based goals that are represented by different attack paths\cite{10.1145/2898375.2898390}. To create accurate and effective attack paths for ICS environments, the use of taxonomies and ATT\&CK characteristics defined by MITRE\cite{9486331} have been employed. ATT\&CK by MITRE is primarily a knowledge library of adversarial strategies as well as a breakdown and classification of offensively focused activities that may be utilized against multiple systems. It focuses more on how attackers interact with systems throughout an operation than on the tools and viruses they deploy. \\
\par ATT\&CK puts these strategies together into a set of tactics to help explain and put the technique in its proper context. Tactics are important contextual categories for particular approaches since they encompass conventional, higher-level notations for what attackers do during an operation, such as persist, find information, move laterally, execute files, and exfiltrate data. The ATT\&CK Matrix represented in Table~\ref{Mitre_table} and Table~\ref{Mitre_table1} illustrates the link between tactics and methods. For example, the strategy of persistence (the adversary's purpose to persist in the target environment) includes a number of approaches such as AppInit DLLs, New Services, and Scheduled Tasks. Each of them is a single tactic that opponents may employ to attain the aim of persistence.  \\
\par As it provides a matrix with multiple possible actions that can be taken by the attacker at each stage in the attack process, enabling us to assess a challenging multi-level attack. It is of particular interest to us as it is the latest standard for industrial control systems and has been proven to be practical in real-world attacks. Understanding the system and its security problems makes attack paths easier to adopt and comprehend. This strategy has been employed with STRIDE and other threat modeling methods in recent years which presents a much better understanding of the modeled threats.

\begin{sidewaystable}[!htpb]
\caption{Tactics and Techniques Representing The MITRE ATT\&CK\textsuperscript{\textregistered} Matrix for Industrial Control Systems (ICSs) (Continued)~\cite{9486331}}
\label{Mitre_table}
    \newcolumntype{C}{>{\centering\arraybackslash}X}
    \begin{tabularx}{\linewidth}{p{2cm}*{12}{C}}
        \toprule
        \textbf{Initial Access} & \textbf{Execution} & \textbf{Persistence} & \textbf{Privilege Escalation} & \textbf{Evasion} & \textbf{Discovery} & \textbf{Lateral Movement} & \textbf{Collection} & \textbf{Command and Control} & \textbf{Inhibit Response Function} & \textbf{Impair Process Control} & \textbf{Impact} \\
        \midrule
        12 techniques & 9 techniques & 6 techniques & 2 techniques & 6 techniques & 5 techniques & 7 techniques & 11 techniques & 3 techniques & 14 techniques & 5 techniques & 12 techniques \\
        \midrule
        Drive-by Compromise & Change Operating Mode & Hardcoded Credentials & Exploitation for Privilege Escalation & Change Operating Mode & Network Connection Enumeration & Default Credentials & Adversary in the Middle & Commonly Used Port & Activate Firmware Update Mode & Brute Force I/O & Damage to Property \\
        \midrule
        Exploit Public-Facing Application & CommandLine Interface & Modify Program & Hooking & Exploitation for Evasion & Network Sniffing & Exploitation of Remote Services & Automated Collection & Connection Proxy & Alarm Suppression & Modify Parameter & Denial of Control \\
        \midrule
        Exploitation of Remote Services & Execution through API & Module Firmware & & Indicator Removal on Host & Remote System Discovery & Hardcoded Credentials & Data from Information Repositories & Standard Application Layer Protocol & Block Command Message & Module Firmware & Denial of View \\
        \midrule
        External Remote Services & Graphical User Interface & Project File Infection & & Masquerading & Remote System Information Discovery & Lateral Tool Transfer & Data from Local System & & Block Reporting Message & Spoof Reporting Message & Loss of Availability \\
        \midrule
        Internet Accessible Device & Hooking & System Firmware & & Rootkit & Wireless Sniffing & Program Download & Detect Operating Mode & & Block Serial COM & Unauthorized Command Message & Loss of Control \\
        \midrule
        Remote Services & Modify Controller Tasking & Valid Accounts & & Spoof Reporting Message & & Remote Services & I/O Image & & Change Credential & & Loss of Productivity and Revenue \\
        \midrule
        Replication Through Removable Media & Native API & & & & & Valid Accounts & Monitor Process State & & Data Destruction & & Loss of Protection \\
        \midrule
        Rogue Master & Scripting & & & & & & Point \& Tag Identification & & Denial of Service & & Loss of Safety \\
        \bottomrule
    \end{tabularx}
\end{sidewaystable}

\begin{sidewaystable}[!htpb]
\caption{Tactics and Techniques Representing The MITRE ATT\&CK\textsuperscript{\textregistered} Matrix for Industrial Control Systems (ICSs) (Continued)~\cite{9486331}}
\label{Mitre_table1}
    \newcolumntype{C}{>{\centering\arraybackslash}X}
    \begin{tabularx}{\linewidth}{p{2cm}*{12}{C}}
    
        \toprule
        \textbf{Initial Access} & \textbf{Execution} & \textbf{Persistence} & \textbf{Privilege Escalation} & \textbf{Evasion} & \textbf{Discovery} & \textbf{Lateral Movement} & \textbf{Collection} & \textbf{Command and Control} & \textbf{Inhibit Response Function} & \textbf{Impair Process Control} & \textbf{Impact} \\
        \midrule
        Spearphishing Attachment & User Execution & & & & & & Program Upload & & Device Restart/Shutdown & & Loss of View \\
        \midrule
        Supply Chain Compromise & & & & & & & Screen Capture & & Manipulate I/O Image & & Manipulation of Control \\
        \midrule
        Transient Cyber Asset & & & & & & & Wireless Sniffing & & Modify Alarm Settings & & Manipulation of View \\
        \midrule
        Wireless Compromise & & & & & & & & & Rootkit & & Theft of Operational Information \\
        \midrule
        & & & & & & & & & Service Stop & & \\
        \midrule
        & & & & & & & & & System Firmware & & \\
        \bottomrule
    \end{tabularx}
\end{sidewaystable}

\subsection{Vulnerability Scoring}
 The vulnerability score measurement is done using the Common Vulnerability Scoring System (CVSS) at step 6 of the process. Since scoring with CVSS requires inputs regarding the specifics of attack vectors and therefore it is positioned at the end of the process.

This score acts as a prioritization mechanism for security teams in the organization. This information is also crucial for Chief Information Security Officers (CISO) and Chief Security Officers so that they can be equipped to mitigate the threats assessed through the process.

Security vulnerability severity is assessed and communicated using the CVSS. CVSS scores help organizations prioritize and fix vulnerabilities. Base, Temporal, and Environmental CVSS measurements exist~\cite{CVSS-Common_Vulnerability_Scoring_System}.

\begin{figure*}[h]
\centerline{\includegraphics[width=.9\textwidth]{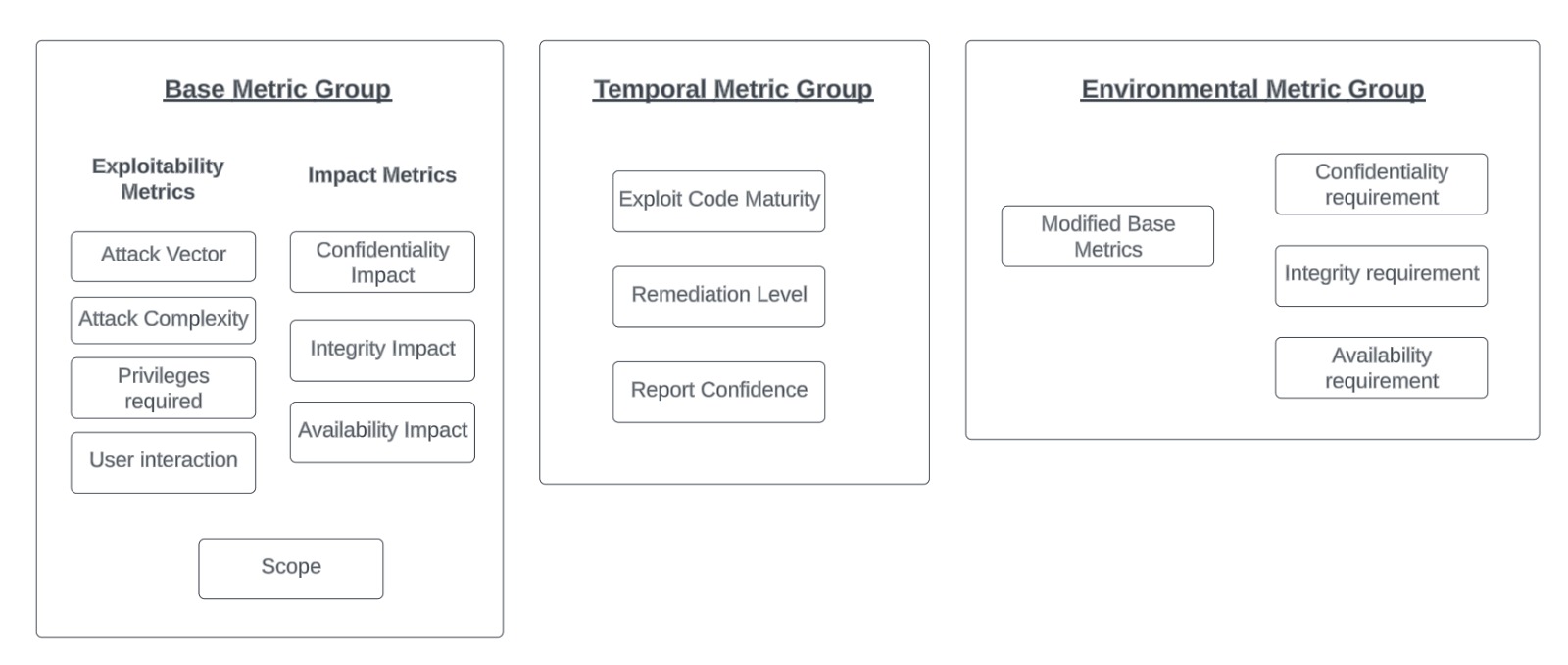}}
    \caption{CVSS Metrics}
\label{cvss_rating}
\end{figure*}
\begin{enumerate}
\item \textbf{Base Metrics:}
These metrics capture the intrinsic qualities of a vulnerability and do not change over time or in different environments. They are used as the foundation for calculating the overall CVSS score. Generally, base metrics include: 
\begin{enumerate}
  
  \item \textbf{Access Vector (AV):} Describes how the vulnerability is accessed. It can be Local, Adjacent Network, or Network.

   \item \textbf{Access Complexity (AC):} Reflects the level of complexity required to exploit the vulnerability. It can be Low, Medium, or High.

  \item \textbf{Authentication (Au):} Indicates the level of authentication required to exploit the vulnerability. It can be None, Single, or Multiple.
  
  \item \textbf{Confidentiality (C), Integrity (I), and Availability (A):} Represent the impact of the vulnerability on these three security attributes. They can be None(0), Partial(0.5), or Complete(1.0).
  
  \item \textbf{Exploitability (E):} Measures how likely it is that the vulnerability will be exploited. It can be Unproven(0.85), Proof-of-Concept(0.9), Functional(0.95), or High(1.0).
  
  \item \textbf{Impact (I):} Reflects the overall impact on the system if the vulnerability is exploited. It can be None(0), Low(0.22), Medium(0.56), High(0.7), or Critical(0.85).

\end{enumerate}
\textit{AV = Network, AC = Low, Au = Single, C = Complete, I = Complete, A = Complete, E = Functional, I = High}

\item \textbf{Temporal Metrics:}
These metrics reflect the characteristics of a vulnerability that change over time but are not specific to a particular environment. They provide a dynamic view of the vulnerability's risk. Generally, temporal metrics include:
\begin{enumerate}
  \item \textbf{Exploit Code Maturity (E):} Indicates the current state of exploit development. It can be Not Defined(0.0), Unproven(0.9), Proof-of-Concept(0.95), Functional(1.0), or High(1.0).
  \item \textbf{Remediation Level (RL):} Represents the official response to the vulnerability. It can be Official Fix(0.0), Temporary Fix(0.25), Workaround(0.75), or Unavailable(1.0).
  \item \textbf{Report Confidence (RC):} Reflects the confidence in the accuracy of the reported vulnerability information. It can be Unconfirmed(0.0), Uncorroborated(0.5), or Confirmed(1.0).
\end{enumerate}
\textit{E = Functional, RL = Official Fix, RC = Confirmed}

\item \textbf{Environmental Metrics:}
These metrics consider the unique characteristics of the environment in which the vulnerability exists. They allow organizations to tailor the CVSS score to their specific circumstances. Generally, environmental metrics include:
\begin{enumerate}
  \item \textbf{Collateral Damage Potential (CDP):} Measures the potential impact on systems beyond the one directly affected. It can be None(0.0), Low(0.1), Low-Medium(0.3), Medium-High(0.4), or High(0.5).
  \item \textbf{Target Distribution (TD):} Reflects the proportion of systems in the environment that are affected. It can be None(0.0), Low(0.25), Medium(0.75), or High(1.0).
  \item \textbf{Confidentiality Requirement (CR), Integrity Requirement (IR), and Availability Requirement (AR):} Represent the importance of security attributes in the environment. They can be Not Defined(0.0), Low(0.5), Medium(1.0), or High(1.51).
\end{enumerate}
\textit{CDP = Low, TD = Medium, CR = High, IR = Medium, AR = Low}

To calculate the overall CVSS score, we can use the following formula as :
\begin{equation}
\text{CVSS Score} = \left( \frac{{(1 - (1 - C) \cdot (1 - I) \cdot (1 - A)) \cdot (E \cdot RL \cdot RC)}}{{(0.6 \cdot (1 - C) + 0.4 \cdot (1 - CDP)) \cdot \text{Impact} + 0.6 \cdot CDP}} \right) \cdot \text{Exploitability}
\end{equation}
\end{enumerate}
\section{Case studies}
\subsection{Electric Power Systems (IoP)}
An electric power system refers to a complex network of electrical components that are designed to facilitate the generation, transmission, and distribution of electricity for various applications. The power system, exemplified by the electrical grid, serves as a means of distributing electricity to both residential and commercial consumers across a vast geographical area. The electrical grid comprises three primary components, namely the power generators responsible for electricity production, the transmission system that facilitates the transportation of power from the production centers to the load centers, and the distribution system that disseminates the power to the neighboring households and businesses. 

The electric power industry is widely acknowledged as a crucial sector, given that its impairment or annihilation would result in a severe impact on security, national economic stability, public health, and safety, or a combination of these factors. Consequently, the process of threat modeling holds significant importance for the aforementioned industry as it aids in the identification of potential threats aimed at a specific target and provides a rationale for the implementation of security measures. And, for this study, industry standards of the electric power industry from ~\cite{SAYED2017481} have been we have considered.\newline

\begin{figure*}[!htp]
\centering
    {\includegraphics[width=1\textwidth]{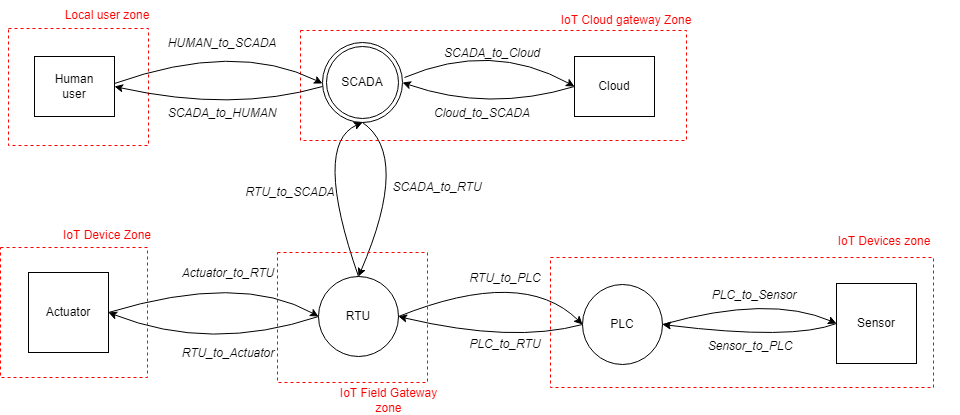}}
    \caption{DFD for SCADA in Electric power system}
    \label{DFD SCA}
\end{figure*}

\begin{figure}[h]
  \centering
  \includegraphics[width=0.7\textwidth]{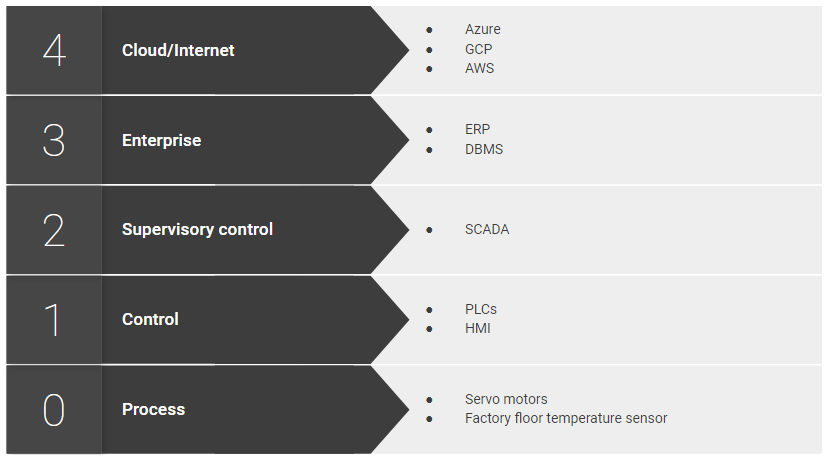}
  \caption{Purdue model for Electric power systems}
  \label{PUR_SCA}
\end{figure}

\textbf{Threat modeling}: The first step in threat modeling is architecture setup using the Purdue model. Fig.~\ref{PUR_SCA} shows the Purdue model for electric power systems. For this analysis, the Supervisory Control and Data Acquisition (SCADA) systems have been chosen. SCADA is placed in the grid control center, which has under its control many remote substations and gathers data from them using the Remote Terminal Unit (RTU). SCADA enables human managers to operate these substations remotely. It also compiles the data being gathered in real time to generate backlogs and provide insights about the safety of devices.

Fig.~\ref{DFD SCA} shows the data flow diagram for a SCADA device in the electric power system. Using the data flows, the possible attack vectors were derived using the STRIDE modeling technique with Microsoft Threat Modeling Tool\cite{jegeib}, through which we can find the number of modeled attacks and threat counts per asset mentioned in DFD\footnote{Computation using the Microsoft Threat Modelling Tool (MTMT)\cite{jegeib} was done on a computer running Windows 10 Pro with an Intel 9300H CPU and an NVIDIA GTX 1650 GPU.
}. The devices denoted with circles are considered IoT gateways, meaning they interface with multiple IoT devices to gather or pass data, and rectangles represent IoT devices. Table ~\ref{tab:2} gives a summary of the number of modeled attacks on electrical power systems.

\begin{figure*}[h]
\centering
\includegraphics[width=1\textwidth]{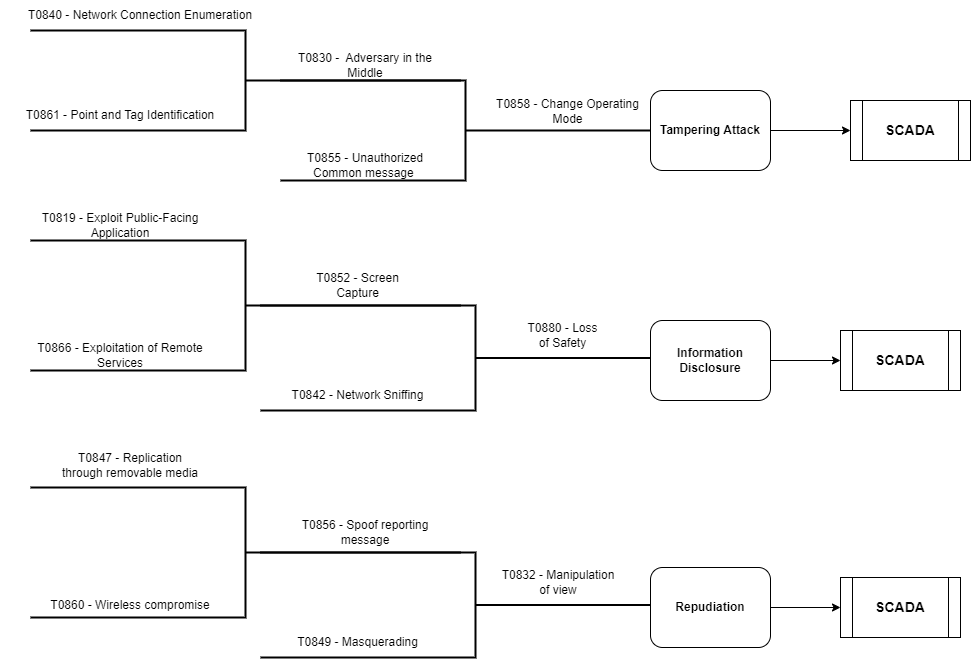}
\caption{Attack paths for SCADA in Electric power system}
\label{AT SCA}
\end{figure*}

\begin{table}[!htp]\centering
\caption{Threat summary for Electric power systems}\label{tab:2}
\scriptsize
\begin{tabular}{lrr}\toprule
Category &Number of modeled attacks \\\midrule
Denial Of Service &20 \\
Elevation Of Privilege &45 \\
Information Disclosure &12 \\
Repudiation &22 \\
Spoofing &26 \\
Tampering &47 \\
\bottomrule
\end{tabular}
\end{table}
After gathering all the attacks, attack paths could be drawn using the MITRE\cite{9486331} notation. Attack paths enable us to understand the flow of the attack from origin to destination and the CWEs exploited in the process. Fig.~\ref{AT SCA} shows the attack paths for 3 such attacks modeled for SCADA.\\

\textbf{CVSS}: As discussed above, the CVSS is an industry-standard scoring system for detected vulnerabilities. In this case study, the same standard from the NVD~\cite{NVD-search_and_statistics} has been utilized to quantitatively rank the vulnerabilities in the order in which they need to be prioritized. The number of attacks for each device is calculated and tabulated in Table~\ref{tab:3}.

\begin{table}[h]\centering
\label{cnt ASSET}
\caption{Threat count per asset}\label{tab:3}
\scriptsize
\begin{tabular}{lrr}\toprule
Asset &Threat count \\\midrule
SCADA &41 \\
Cloud &13 \\
PLC &36 \\
RTU &57 \\
Actuator &12 \\
Sensor &9 \\
\bottomrule
\end{tabular}
\end{table}

CVSS scores can be calculated using the available CVSS calculator. This calculation must be precise and done by experts. So, to get the reality of CVSS scores, consider the example of Schneider Electric, whose vulnerabilities are scored and are available at the NVD~\cite{NVD-search_and_statistics}. Schneider Electric SE is a European corporation that has expertise in the realm of energy management. Its focus is on providing solutions for a variety of sectors, including homes, buildings, data centers, infrastructure, and industries. Table~\ref{tab:4} shows the top 5 threats identified by using this method from the NVD.

\begin{table}[!htp]\centering
\caption{Top 5 threats}
\label{tab:4}
\scriptsize
\begin{tabular}{lrrrr}\toprule
Category &Interaction &CVSS Score \\\midrule
Elevation Of Privilege &plc\_to\_rtu &9.8 \\
Spoofing &human\_to\_plc &9.1 \\
Spoofing &human\_to\_scada &8.1 \\
Tampering  &human\_to\_scada &7.5 \\
Denial Of Service &rtu\_to\_plc &7.5 \\
\bottomrule
\end{tabular}
\end{table}

\subsection{Manufacturing Industry (IoM)}
The manufacturing industry comprises establishments engaged in the mechanical, physical, or chemical transformation of materials, substances, or components into new products. Examples of manufacturing companies include automotive companies, petroleum-based companies, food production, and processing companies, etc.
\par The manufacturing industry is widely seen as a crucial sector due to its potential to significantly impact security, national economic security, national public health, or safety, either alone or in combination. Therefore, the practice of threat modeling is of significant importance in this industry for identifying a list of threats against a target and justifying security efforts.
\par
\begin{figure*}[h]
  \centering
\includegraphics[width=1\textwidth]{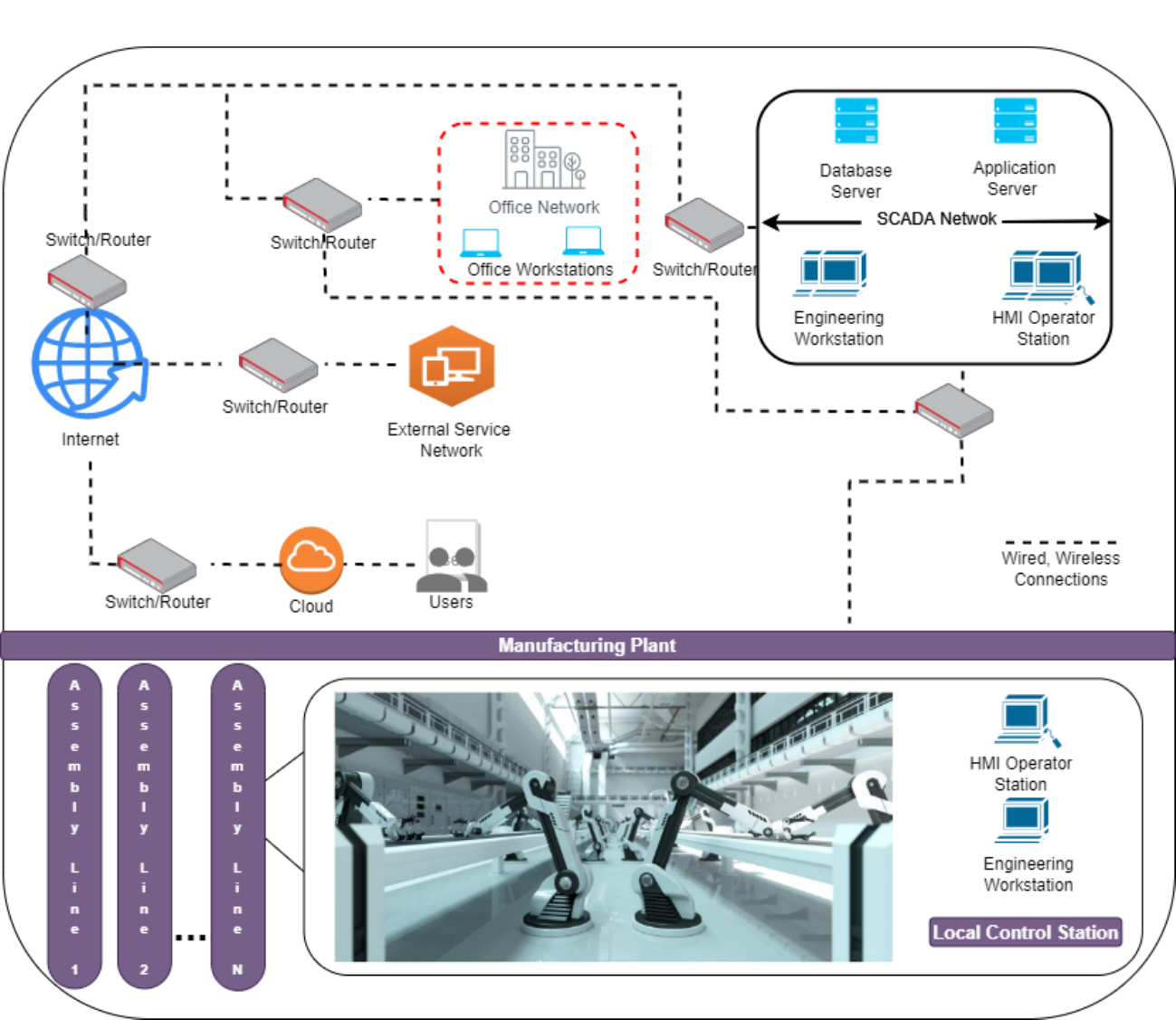}
  \caption{The Architecture of A Manufacturing Plant}
  \label{architecture}
\end{figure*}

 In this section, for the purpose of security analysis, human resources such as operators, contractors, and engineers have been excluded as the scope of this study did not cover how the operational security of a smart manufacturing system is influenced by human behavior (Fig.~\ref{architecture}). This document delineates the segregation of the factory floor network from other networks, such as the Internet and corporate network, via the use of firewalls, here the red nodes may be utilized as access points for attacks~\cite{JBAIR2022103611}.

\begin{figure*}[h]
\centering
\includegraphics[width=1\textwidth]{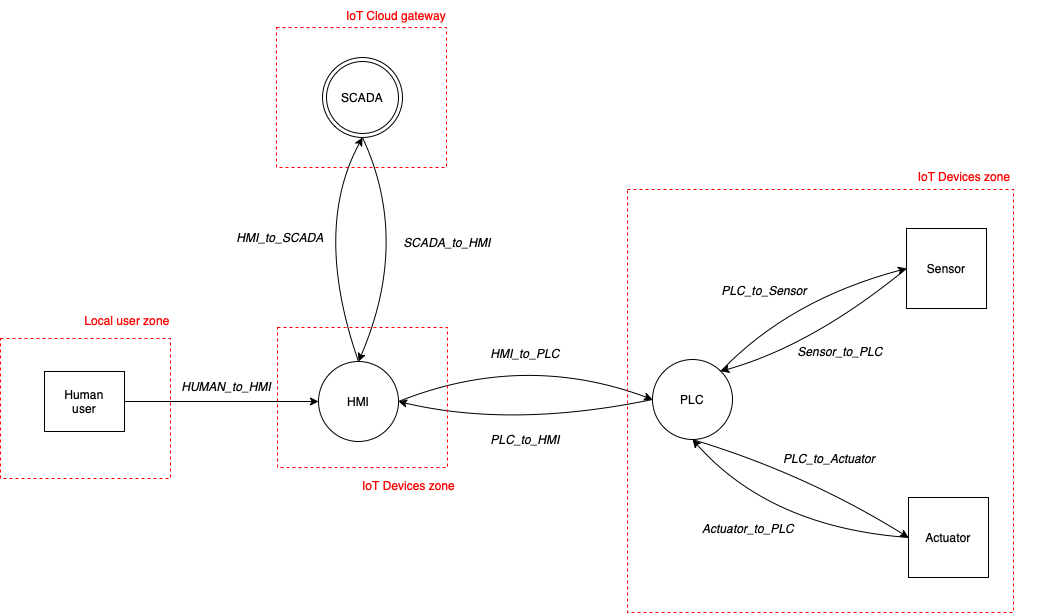}
\caption{DFD for HMI in a Manufacturing Industry}
\label{DFD HMI}
\end{figure*}

\begin{figure*}[t!]
\centering
\captionsetup{justification=centering, margin=2cm}
\centering
\includegraphics[width=1\textwidth]{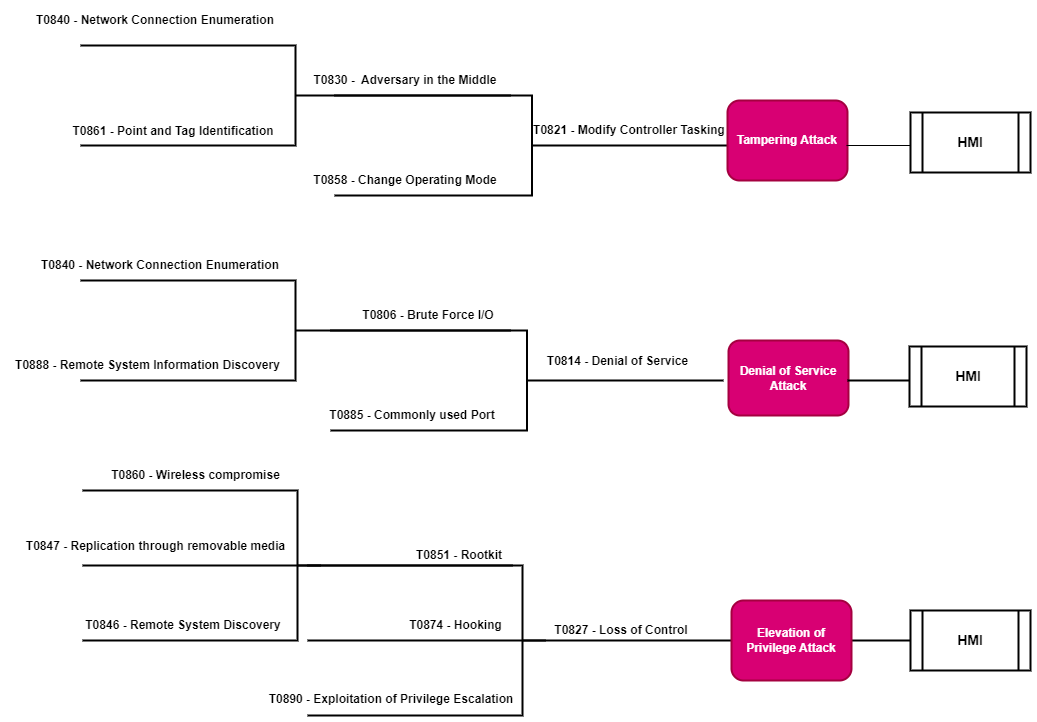}
\caption{Attack paths for HMI in a Manufacturing Industry}
\label{AT HMI}
\end{figure*}

\begin{figure}[htp!]
  \centering
  \includegraphics[width=0.8\textwidth]{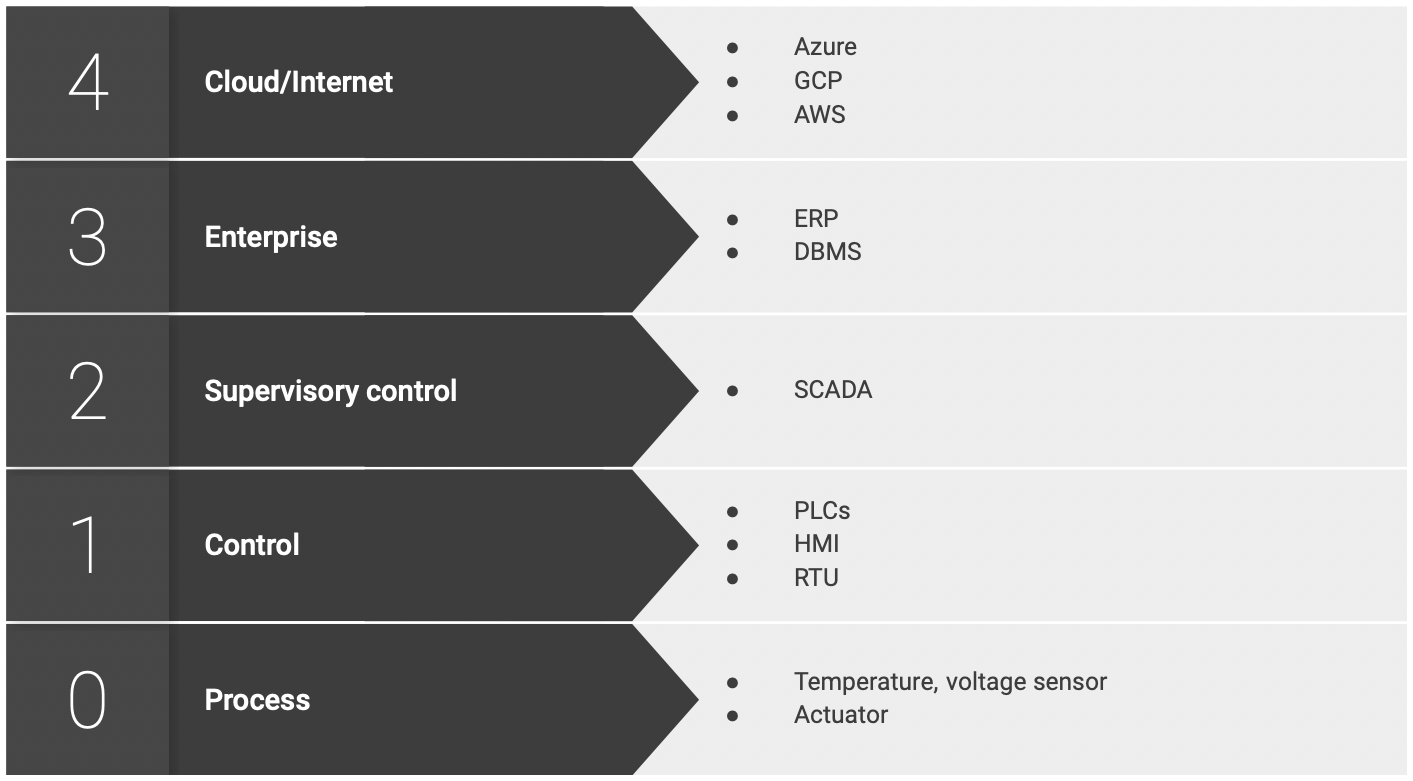}
  \caption{Purdue model for a Manufacturing Industry}
  \label{PURDUE HMI}
\end{figure}
\textbf{Threat modeling}: The first step in threat modeling is asset selection using the Purdue model. Fig.~\ref{PURDUE HMI} shows the Purdue model for the manufacturing industry. The Human Machine Interface (HMI) has been used for this analysis. HMI is a software and hardware combination that allows the operator to interact between systems and machines. The HMI system is unavoidable; in most manufacturing industries, the automation system would be incomplete without an HMI.\\ 
An HMI allows an operator to monitor the floor's activity in real time. If there is a problem, the HMI will notify the operator. In essence, the HMI serves as the user interface in a manufacturing or process control system. The HMI also provides a graphical representation of the industrial control and monitoring system. 

Fig.~\ref{DFD HMI} shows the data flow diagram for an HMI device in the manufacturing industry. Similar to the Electric power system case study, by using the data flows, all possible attack vectors could be drawn using the STRIDE modeling technique through the Microsoft Threat Modelling Tool\cite{jegeib}. The devices denoted with circles are considered IoT gateways, meaning they interface with multiple IoT devices to gather or pass data, and rectangles represent IoT devices. Table~\ref{tab:5} gives a summary of the number of modeled attacks on the manufacturing industry.

\begin{table}[!htp]\centering
\caption{Threat summary for Manufacturing Industry}\label{tab:5}
\scriptsize
\begin{tabular}{lrr}\toprule
Category &Number of modeled attacks \\\midrule
Denial Of Service &10 \\
Elevation Of Privilege &32 \\
Information Disclosure &9 \\
Repudiation &9 \\
Spoofing &11 \\
Tampering &39 \\
\bottomrule
\end{tabular}
\end{table}

After gathering all the attacks, attack paths could be drawn using the MITRE \cite{9486331} notation. Attack paths enable us to understand the flow of the attack from origin to destination and the CWEs exploited in the process. Fig.~\ref{AT HMI} shows the attack paths for 3 such attacks modeled for HMI.\newline

\textbf{CVSS}: Similar to the first case study, here the same standard from the NVD~\cite{NVD-search_and_statistics} has been utilized to quantitatively rank the vulnerabilities in the order in which they need to be prioritized. 
The number of attacks for each device is calculated and tabulated in Table~\ref{tab:6}.

\begin{table}[!htp]\centering
\caption{Threat count per asset}\label{tab:6}
\scriptsize
\begin{tabular}{lrr}\toprule
Asset&Threat count \\\midrule
PLC &47\\ 
SCADA &11 \\ 
HMI &34 \\ 
Actuator &9 \\ 
Sensor &9 \\
\bottomrule
\end{tabular}
\end{table}

CVSS scores can be calculated using the available CVSS calculator. This calculation must be precise and done by experts. So, to get the reality of CVSS scores, the example of Johnson \& Johnson is considered, whose vulnerabilities are scored and are available at NVD. Johnson \& Johnson is a large American multinational corporation that specializes in the development, manufacturing, and distribution of a broad range of healthcare products. Johnson \& Johnson's products include medical devices, pharmaceuticals, and consumer healthcare products such as Band-Aids, Tylenol, and Neutrogena. Table~\ref{tab:7} shows the top 5 threats identified by using this method from the NVD~\cite{NVD-search_and_statistics}.

\begin{table}[!htp]
\centering
\caption{Top 5 threats}\label{tab:7}
\scriptsize
\begin{tabular}{lrrrr}\toprule
Category& Interaction & CVSS Score 
\\\midrule
Spoofing  & sensor\_to\_plc & 9.8 \\
Tampering  & plc\_to\_sensor & 9.8 \\
Spoofing  & actuator\_to\_plc & 7.5 \\
Information Disclosure & plc\_to\_hmi & 7.5 \\
Spoofing  &plc\_to\_hmi & 4.8\\
\bottomrule
\end{tabular}
\end{table}

\section{Conclusion}
In order to ensure the deployment of a secure Industrial Control and Production System (ICPS) in smart factories, it is crucial to address potential cyber threats and attacks, and to understand the associated risks. This study proposes a comprehensive threat modeling and attack path analysis framework for industrial IoT systems that have wide-ranging applicability and uses standardized methods to deploy an effective solution in six steps. 
The proposed work identifies possible threats that are based on STRIDE modeling and attack paths with vulnerability scores derived from CVSS. In contrast to the existing research gap, there is a plethora of research on threat modeling and detection, there is a conspicuous absence of a standardized, step-by-step process in the literature for identifying, assessing, and prioritizing threats, especially in the context of Industrial IoT also, none of the models and methodologies presented in the findings are tailored for specific industries. While these offer valuable insights for particular sectors, their applicability becomes limited when faced with the diverse and evolving landscape of the Internet of Things across various industries. 
\par  The proposed work fills this gap by presenting a systematic approach and offers a structured methodology. This approach not only aids in the identification of threats but also offers metrics and criteria for their assessment and prioritization. Such a methodology is vital for organizations to address the most critical vulnerabilities first, ensuring efficient allocation of resources and timely mitigation. One of the salient features of our work is its industry-agnostic nature. The design is centered on being versatile and adaptable, ensuring that it remains relevant across a wide spectrum of industries, from manufacturing to production. This broad applicability ensures that this method can serve as a foundational tool for organizations irrespective of their domain, enhancing its utility and impact. To highlight the potential benefits for researchers, practitioners, and decision-makers in understanding cyber security for ICPS and guiding investments in this field, two case studies, namely the industrial manufacturing line (IoM) and the power industry (IoP) are also discussed in this paper. The proposed approach of threat modeling and attack path analysis provides a clear picture of the threats in the cyber-physical landscape of industrial IoT. This helps in designing systems that can be better protected to handle the cyber threats of today and mitigate the effects of a potential compromise on the system. \\

In conclusion, this study underscores the critical need for a robust approach to address cyber threats and security in the realm of Industrial Control and Production Systems (ICPS) deployed in smart factories. The proposed comprehensive framework for threat modeling and attack path analysis, consisting of six systematic steps, represents a significant contribution to the field of industrial IoT. While existing research has made substantial strides in threat modeling and detection, a noticeable gap exists in the absence of a standardized, step-by-step process for identifying, assessing, and prioritizing threats, particularly within the context of Industrial IoT.

Moreover, the study highlights the lack of tailored models and methodologies for specific industries, limiting their applicability in a rapidly evolving landscape. The proposed approach not only fills this gap but also introduces a structured methodology, complete with metrics and criteria for assessing and prioritizing threats. Such a methodology is of paramount importance for organizations aiming to efficiently allocate resources and ensure timely threat mitigation.

An important aspect of this work is its industry-agnostic nature, designed to be versatile and adaptable, making it relevant across a wide spectrum of industries, from manufacturing to production. This versatility positions the method as a foundational tool for organizations, regardless of their domain, ultimately enhancing its utility and impact.

Additionally, the study showcases the practical benefits of this research through two case studies in the industrial manufacturing line and the power industry. By adopting the proposed approach of threat modeling and attack path analysis, organizations can gain a clearer understanding of the threats in the cyber-physical landscape of industrial IoT. This, in turn, empowers them to design systems that are better protected against contemporary cyber threats, mitigating the potential impacts of system compromise.

\section{Limitations and future work}
 The paper also suggests several areas for future work, including the development of models to predict financial losses and business performance due to cyber threats. The work on risk assessment may also be targeted as a future scope of this work.

\section{Acknowledgements}
The authors are thankful to Data Science \& Business Informatics (DSBI), IIIT Allahabad group members for encouraging this work especially Dr. Nachiket Tapas and Mrs. Uphar Singh for their suggestions that greatly improved the manuscript.
\\
\textbf{Declaration of Competing Interests:} The authors affirm that they own no identifiable financial conflicts of interest or personal affiliations that might have potentially influenced the findings presented in this research article.
\\
\textbf{Authors Contribution Statement:}
Each of the authors made significant contributions to the idea and design of the research. The tasks of material preparation, data collecting, and analysis were carried out by Kumar Saurabh, Deepak Gajjala, and Krishna
Kaipa. The first draft of
the manuscript was reviewed and supervised by Rahamatullah Khondoker, Ranjana Vyas, and O.P. Vyas. The final document was reviewed and approved by all of the authors.
\\
\textbf{Funding:} The study conducted did not receive any financial support from external sources except the doctoral fellowship for research from the Indian Institute of Information Technology Allahabad supported by the Ministry of Education (MoE), India.
\\
\textbf{Data Availability and Access:} Data sharing is not relevant in this research since there were no datasets created or analyzed. This paper does not have any linked data sets.

\bibliography{sn-bibliography.bib}
\end{document}